\documentclass[aps,prl,amsmath,superscriptaddress,amssymb,reprint,floatfix]{revtex4-2}
\usepackage{subfiles}
\usepackage{physics}
\usepackage{graphicx}% Include figure files
\usepackage{graphics}
\usepackage{booktabs} % configure tables
\usepackage[colorlinks=true,linkcolor=red]{hyperref}   % use for hypertext links, including those to external documents and URLs
\usepackage{placeins} %Use the \FloatBarrier command before the conclusion section to ensure floats don’t go past it.
\usepackage{dcolumn}% Align table columns on decimal point
\usepackage{bm}% bold math
\usepackage[caption=false]{subfig}
\usepackage{algorithm}
\usepackage{algpseudocode}
\usepackage{amsmath}
\begin{document}

\title{Machine Learning Assisted Sorting of Active Microswimmers}

\author{Abdolhalim Torrik}
\email{yashar.torrik@gmail.com}
\affiliation{Department of Physical and Computational Chemistry, Shahid Beheshti University, Tehran 19839-9411, Iran} 

\author{Mahdi Zarif}
\email{m\_zarif@sbu.ac.ir}\thanks{Corresponding author: m\_zarif@sbu.ac.ir}
\affiliation{Department of Physical and Computational Chemistry, Shahid Beheshti University, Tehran 19839-9411, Iran}

\date{\today}

\begin{abstract}

Active matter systems being in a non-equilibrium state, exhibit complex behaviors such as self-organization and giving rise to emergent phenomena. There are many examples of active particles with biological origins, including bacteria and spermatozoa, or with artificial origins, such as self-propelled swimmers and Janus particles. The ability to manipulate active particles is vital for their effective application e.g. separating motile spermatozoa from nonmotile and dead ones, to increase fertilization chance. In this study, we proposed a mechanism -- an apparatus -- to sort and demix active particles based on their motility values (P\`eclet number). Initially, using Brownian simulations we demonstrated the feasibility of sorting self-propelled particles. Following this, we employed machine learning methods, supplemented with data from comprehensive simulations that we conducted for this study, to model the complex behavior of active particles. This enabled us to sort them based on their P\`eclet number. Finally, we evaluated the performance of the developed models and showed their effectiveness in demixing and sorting the active particles. Our findings can find applications in various fields, including physics, biology, and biomedical science, where the sorting and manipulation of active particles play a pivotal role.
\end{abstract}
\date{\today}

%\pacs{ 64.70. Pf}

\maketitle
%%--------------------Intro----------------------------------------------------------
\section{INTRODUCTION}
\label{sec:Intro}
	
	Active matter systems -- living and non-living -- are defined by their ability to consume energy, generate forces, and show synchronized self-organization \cite{das2020introduction, volpe2022active}. These systems are intrinsically nonequilibrium and show complex behaviors and emergent phenomena \cite{likos2001effective, romanczuk2012active, bechinger2016active, hagan2016emergent}. They exist in various forms and scales, and can be biological or artificial, macroscopic or microscopic in nature. Examples of biological active systems with macroscopic origin include: flocks of birds \cite{ballerini2008interaction}, schools of fish \cite{moussaid2009collective}, swarms of insects \cite{buhl2006disorder,tennenbaum2016mechanics}, and herds of mammals \cite{vicsek2012collective}. Microscopic examples to highlight the dynamics of biological active matter systems include: bacteria \cite{kearns2010field,berg:book2003}, sperm cells \cite{gaffney2011mammalian,woolley2003motility}, amoeba cell clusters \cite{rappel1999self,kessler1993pattern,nagano1998diffusion}, microbial biofilms \cite{drescher2016architectural}, marine algae \cite{Goldstein:ARFM2015}, and zooplanktons \cite{jekely2008mechanism}. Self-propelled artificial swimmers (catalytic, magnetic and chemically driven micro-motors) \cite{paxton2004catalytic,kumar2015self,takatori2016acoustic,narayan2007long,kokot2017dynamic,zhang2009artificial,sanchez2015chemically,rao2015force,mohanty2020contactless,zhou2021magnetically,sitti2015biomedical}, Janus particles \cite{walther2013janus,jiang2010janus,Zhang:langmuir2017,Poggi:CPS2017}, and robotic swarms \cite{turgut2008self} are examples of artificial active matter systems. All of the preceding examples are only a portion of the spectrum of active matter systems.

	Despite advancements in understanding these systems, there’s still a challenge: the effective sorting of active particles based on their motility values (P\`eclet number). Manipulating active particles is a vital part of their effective application, there are many ways to achieve this, examples of such cases are: separation and rectification of swimming bacteria \cite{wan2008rectification, tailleur2009sedimentation, maggi2013motility}, using microswimmers for separating a mixture of colloids \cite{yang2012using}, aggregation and segregation of microswimmers based on their motilities \cite{mccandlish2012spontaneous, yang2014aggregation}, using the confinement to manipulate microswimmers \cite{fily2014dynamics, berdakin2013influence, reichhardt2013dynamics, volpe2014simulation, torrik2021dimeric}, separation \cite{ai2015chirality} and sorting \cite{chen2015sorting,Mijalkov:SoftMatter2015} of microswimmers based on their chirality, motility \cite{costanzo2014motility} and separating motile spermatozoa \cite{guzick2001sperm, chinnasamy2018guidance}. However, a mechanism that can sort active particles based on their motility values, such as the P\`eclet value, is yet to be developed.
	
	%	Machine Learning 
	Machine learning algorithms learn from the data and establish connections between the data to generate decisions/predictions for previously unobserved data without explicit programming \cite{mehta2019high,glielmo2021unsupervised,jordan2015machine,bishop2006pattern}.
	
	In machine learning, there are three main branches: supervised, unsupervised, and semi-supervised learning \cite{bishop2006pattern}. In supervised learning, algorithms are trained to learn from input data and corresponding output data, input is processed to map it to new data based on expected outputs. This algorithms are used on both regression and classification tasks. Unsupervised learning algorithms are different, learning is done without any output data for the given input data. examples of such algorithms are clustering, PCA, etc. Semi-supervised learning falls between supervised and unsupervised learning, in this algorithm, a combination of small labeled data and a large amount of unlabeled data is used to train the model \cite{van2020survey}.
	
	%	Machine Learning Applications
	Machine learning has applications across many scientific and also practical disciplines. Within the fields of medicine and biology, machine learning example use cases are: drug discovery \cite{dara2022machine,vamathevan2019applications,carracedo2021review,patel2020machine,jimenez2020drug,kolluri2022machine}, neuroscience \cite{kording2018roles}, biomedicine \cite{rajpurkar2022ai,goecks2020machine,woldaregay2019data,artzi2020prediction}, cancer research \cite{koh2022artificial,curtis2012genomic,gao2019deepcc,chiu2019predicting,mckinney2020international}, medical imaging \cite{liu2019comparison,siddiqui2020review,esteva2017dermatologist,ozturk2020automated,mckinney2020international,esteva2021deep,xu2020deep,shi2020review,narin2021automatic,wang2020covid}. Machine learning algorithms are also being used for material design \cite{schmidt2019recent,choudhary2022recent,batra2021emerging,tao2021nanoparticle,hart2021machine,aykol2020machine,debroy2021metallurgy,zhu20213d,prashun2017computationally,boyd2017computational}, polymer science \cite{sha2021machine,reis2021machine,patra2021data,tao2021benchmarking,cencer2022machine,gormley2021machine,kim2021polymer}, crystallography \cite{maffettone2021crystallography,banko2021deep,lee2022powder,zimmermann2023finding}, fluid dynamics \cite{vinuesa2022enhancing,vinuesa2023transformative,yousif2023deep,guastoni2023deep,vignon2023recent,yu2022three,kochkov2021machine,brunton2020machine,kutz2017deep,brunton2021applying,fukami2020assessment}, condensed matter physics \cite{bedolla2020machine,carrasquilla2017machine,ferguson2017machine}. Active mater physics also takes the advantage of machine learning \cite{cichos2020machine,cichos2023artificial,essafri2022designing}, example applications include: detecting various phase transitions using Principal Component Analysis (PCA) \cite{jadrich2018unsupervised_a,jadrich2018unsupervised_b}, active particle control using neural networks \cite{franzl2020active,falk2021learning}, identifying anomalous diffusion using recurrent neural networks \cite{argun2021classification}, forecasting dynamics of active nematics using neural networks \cite{colen2021machine,zhou2021machine,frishman2021learning}, extracting the effective interaction of particles from the system configuration \cite{bag2021interaction}, optimal navigation strategies for active particles \cite{nasiri2023optimal,zou2022gait,liu2023learning,hartl2021microswimmers,el2023steering,caraglio2023learning,borra2022reinforcement,yang2022autonomous,larchenko2021study}, characterizing motility induced regimes \cite{mcdermott2023characterizing,dulaney2021machine}, tracking Janus particles in three dimensions \cite{bailey2022tracking}, and applying deep learning algorithms to plankton ecology \cite{bachimanchi2023deep}.
		
	In their study, Galajda et al. \cite{galajda2007wall}, developed an array of funnel-shaped barriers with wide and narrow openings which concentrated bacterial (E.coli) suspensions with motility on the side with narrow openings. This was because motile bacteria could swim into the funnels through wide openings while their ability to do the reverse through narrow openings was severely limited. Paoluzzi et al. Kumar et al. \cite{kumar2019trapping} explored the trapping phase transition of motile polar rods in the presence of a V-shaped obstacle. They found that when the trap angle of the obstacle was below a certain threshold, a trapping transition would occur, while above the threshold angle, all motile particles would escape. Ribeiro et al. \cite{ribeiro2020trapping} considered the influence of spatially periodic potentials on the trapping and sorting of motile active particles. Using simulations they showed that different diffusion regimes and trapping states occur which was a result of the noise value of the active particles and the density of the system. In their work, Ai et al. \cite{Ai:SoftMatter2018} investigated the mixing and demixing of binary mixtures of polar chiral (clockwise and anticlockwise) active particles with polar velocity alignment. They showed that when chirality difference and the polar velocity alignment of active particles compete with each other (neither is dominant), demixing occurs in which particles with clockwise rotation aggregate into one cluster and particles with anticlockwise chirality aggregate into another cluster. Thus using chirality was a means to demix particles. Miska et al. \cite{misko2023selecting} proposed a method incorporating an acoustofluidic setup for selecting particles based on their motility, where particles with high motility would escape from the acoustic trap. They demonstrated their method using both simulations and also experiments with Janus particles and human sperm and proposed that using this method highly motile sperm could be selected for medically assisted reproduction.
	
	Paoluzzi et al. \cite{paoluzzi2020narrow} investigated the narrow escape time of active particles from circular domains using numerical simulations, they showed that narrow escape time undergoes a crossover between two asymptotic regimes with control parameters being the ratio of persistence length of the active particles and the length scale of the circular domain, they suggested the possibility of sorting active particles based on motility parameters. Based on this, we propose a mechanism -- an apparatus -- to sort (demix) active particles based on their motility. We use Brownian dynamics simulations and also incorporate machine learning methods to model the active particles using the data from an extensive number of comprehensive simulations that we conducted for this study.

%%--------------------Method----------------------------------------------------------
\section{METHODS}
\label{sec:methods}

	We consider a rigid circular apparatus with radius $r_{s}$ which is made up of monomers on its circumference each having a radius of $d_{mono}$. The apparatus is inside a planar square box, which also contains active Brownian particles each with a radius $a$. We also pin the apparatus from its center of mass to the center of the box. Thus it can rotate with angular velocity $\omega$, but prevented from moving. The angular velocity value can be varied in different simulations but has a constant value in each. We study our system in 2D, which is convenient to reduce computational resources while being effective in reproducing the key features of active models. 

We use the over-damped Langevin equations to describe the motion of particles as time ($t$) progresses:

\begin{eqnarray}
	\dot{{\mathbf r}}_i &= &V_s{\mathbf n}_i-\mu_T\frac{\partial  U}{\partial {{\mathbf r}_i}}+\sqrt{2 D_T}\, {\boldsymbol \eta}_i(t),
	\label{Eq:langevin_a}
	\\
	\dot{\theta}_i&=&\omega + \sqrt{2D_R}\, \zeta_i(t),
	\label{Eq:langevin_b}
\end{eqnarray} 
here, ${\mathbf r}_i(t) = (x_i(t), y_i(t))$ is the position of the particle, $V_s$ is the self-propulsion of the particle, and ${\mathbf n}_i=(\cos \theta_i, \sin \theta_i)$ is direction of motion with $\theta_i(t)$ as orientation angles. $\mu_T$ is the translational mobility of particles, $\mathbf{f_i}\equiv -{ \partial U}/{\partial {{\mathbf r}_i}}$ is the force acting on each particle, from interactions between the $i$th particle and all other particles. $D_T = \mu_T k_{\mathrm{B}}T$ is the translational diffusion coefficient of particles ($k_{\mathrm{B}}$ is the Boltzmann constant and $T$ is the absolute temperature). $\omega$ is the angular velocity of particles. $D_R$ is the rotational diffusion coefficient of particles which satisfies the relation $D_R = 3D_T/\sigma^2$ in low Reynold's regime. 
The symbols ${\boldsymbol \eta}_i(t)$ and ${\boldsymbol \zeta}_i(t)$ denote independent translational and rotational noises for each particle. They are characterized by white Gaussian distributions with zero mean, such that $\langle {\eta}_i^\alpha(t) \rangle = \langle \zeta_i(t) \rangle=0$. Additionally, they exhibit two-point correlations given by $\langle {\eta}i^\alpha(t) {\eta}j^\beta(t') \rangle={ij}\delta{\alpha\beta}\delta(t-t')$ and $ \langle \zeta_i(t) \zeta_j(t') \rangle=\delta(t-t')$, where $i$ and $j$ represent active particle labels, and $\alpha$ and $\beta$ refer to the Cartesian coordinates $x$ and $y$.
The steric potentials, denoted as $V({\mathbf r}_{ij})$, govern the interactions among active particles and the inclusions. These potentials are derived from modified versions of the Weeks-Chandler-Andersen potential (WCA):
\begin{equation}
	\!\!\!\!V({\mathbf r_{ij}}) \!=\!
	\left\{\begin{array}{l l}
		\!\!4\epsilon\! \left [\! \left ( \frac{\sigma_{\mathrm{eff}}}{|{\mathbf r_{ij}}|} \right )^{12}\! \!-\!2\! \left ( \frac{\sigma_{\mathrm{eff}}}{|{\mathbf r_{ij}}|} \right )^{6}\!\!+\!1\right ] &: |{\mathbf r_{ij}}| \!\leq\! \sigma_{\mathrm{eff}}, \\ 
		\!\!0 &: |{\mathbf r_{ij}}| \!>\! \sigma_{\mathrm{eff}},
	\end{array}\right.
\end{equation}

The {\em P\'eclet number} (or rescaled self-propulsion strength), $Pe_s$, defined as 
\begin{equation}
	Pe_s=\frac{a V_s}{D_T} = \frac{3 V_s}{4D_R a},  
\end{equation}
where we have used $D_R=3D_T/4a^2$ for no-slip spheres in the low-Reynolds-number (Stokes) regime \cite{happel:book1983}. 

Using Brownian dynamics methods, we discretized the simulations over sufficiently small time steps $\Delta \tilde t$ to solve the equations (\ref{Eq:langevin_a}) and (\ref{Eq:langevin_b}) numerically. 

%%--------------------Results----------------------------------------------------------
\section{Results}
\label{sec:results}  

\subsection{Demixing}
\label{sec:demixing}

\begin{figure}[t!]
		\centering
		\subfloat[Initial configuration of example system]{%
			\includegraphics[width=0.45\linewidth, trim={7cm 7cm 7cm 7cm}, clip]{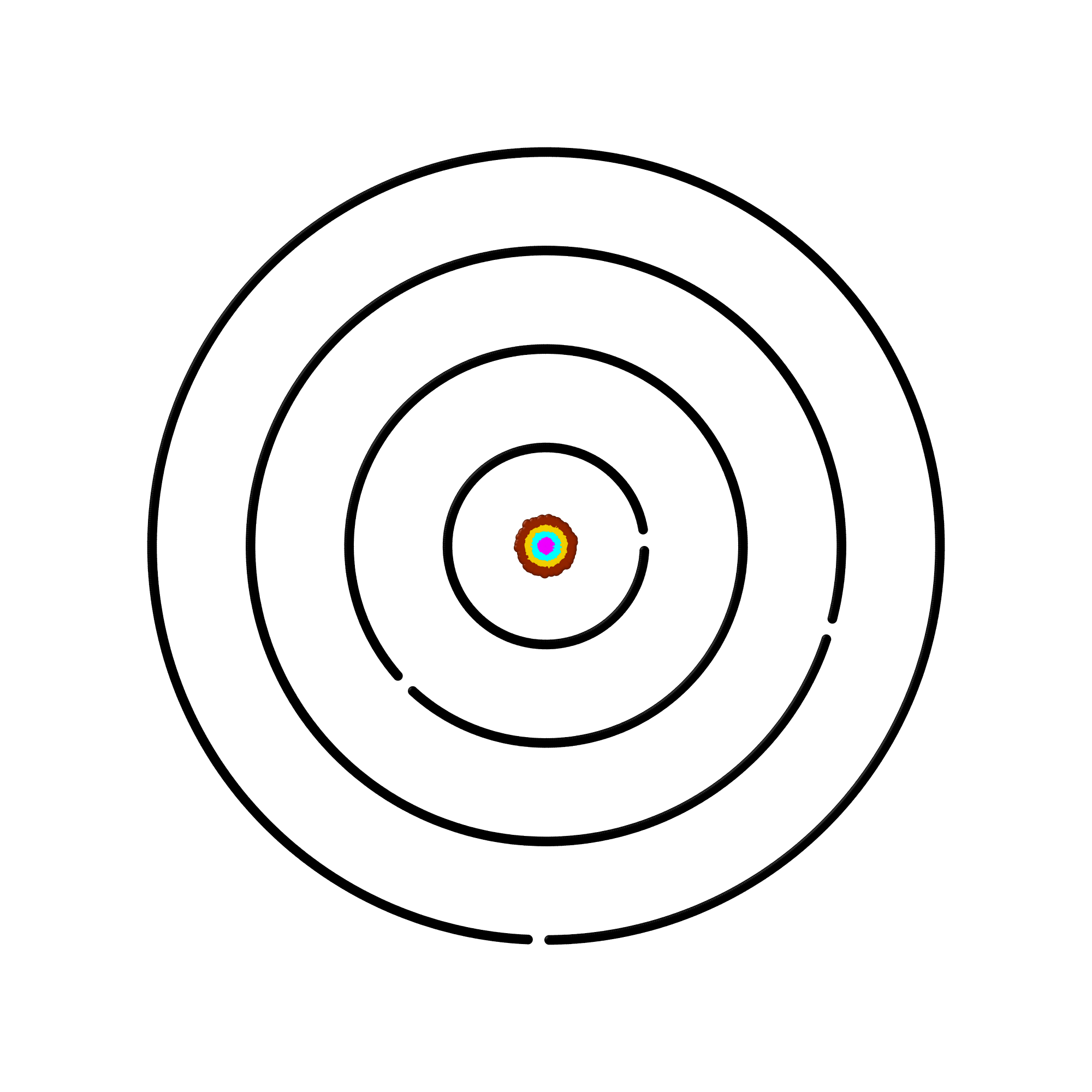}%
			\label{fig:demix-a}%
		}
		\hfill
		\subfloat[Final configuration of example system]{%
			\includegraphics[width=0.45\linewidth, trim={7cm 7cm 7cm 7cm}, clip]{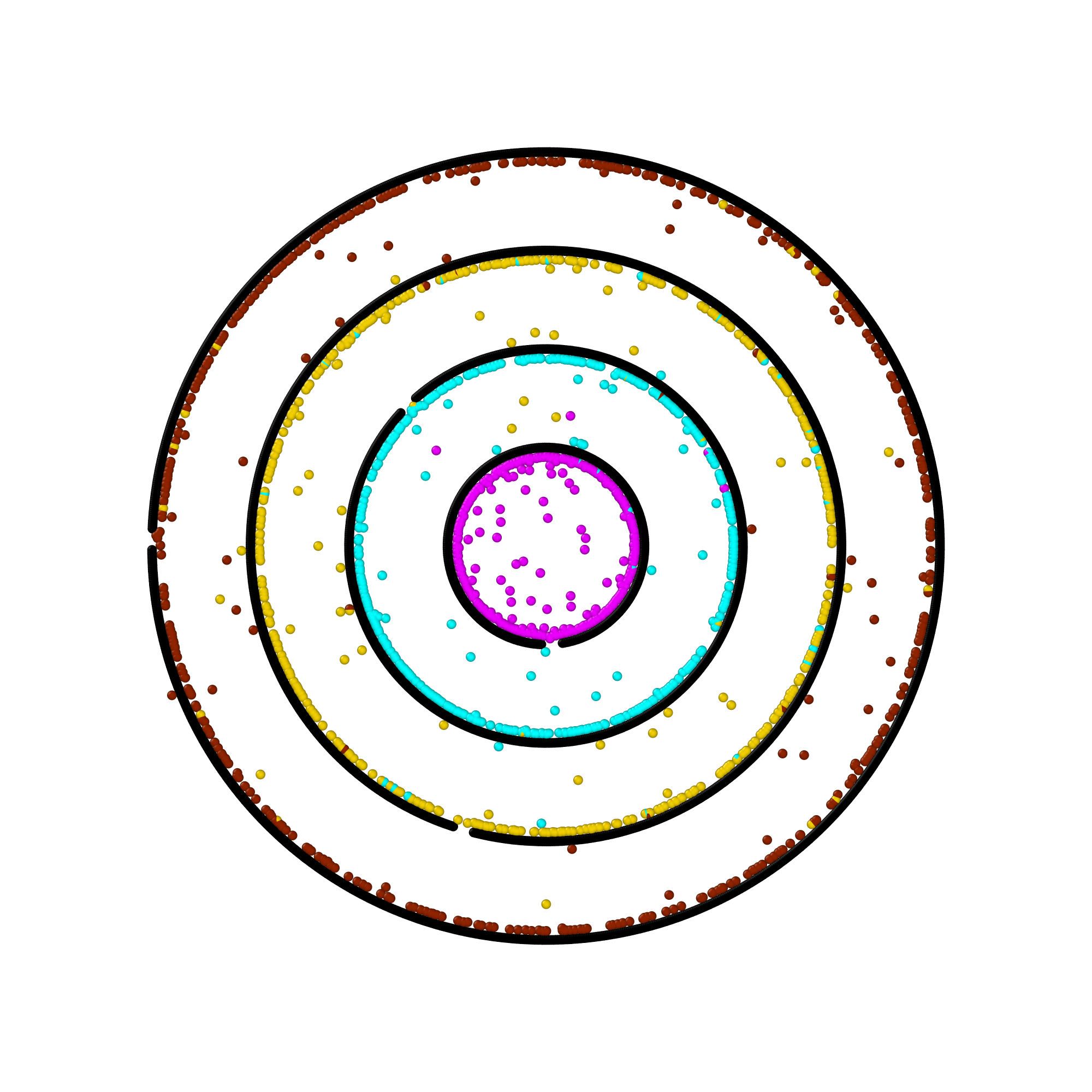}%
			\label{fig:demix-b}%
		}
		\caption{Example of demixing of 4 particles with different $Pe_s$. System defined have $r_s$ = 10, 20, 30 and 40, $d\_mono$ = 1.0, $l_{pore}$ = 2.0, and $k_BT$ = 1.0}
		\label{fig:demix}
	\end{figure}

Fig \ref{fig:demix} shows an example of demixing of four particles with different $Pe_s$, with system having $r_s$ = 10, 20, 30 and 40, with $d_{mono}$ = 1.0, $l_{pore}$ = 2.0, and $k_BT$ = 1.0, with Figs \ref{fig:demix-a} and \ref{fig:demix-b} being the initial and the final configurations, respectively. Demixing of particles in this figure occurs because of the system configuration and $Pe_s$ values of particles that were chosen. Determining the optimal parameters for successful demixing necessitated a process of trial and error. Thus, with the aim of the occurrence of demixing and the ambiguity of the relationship between system parameters and the onset of demixing, we employ a grid search across various parameter combinations and system configurations. By leveraging machine learning methods we aim to gain a better understanding of the demixing phenomenon in active matter systems.
	
To model the demixing of particles from the apparatus, more than 75,000 simulations were done to collect data points with various parameters, namely: $r_s$, $\omega$, $d_{mono}$, $l_{pore}$, and $k_BT$. Data were obtained using Brownian dynamics, simulations aimed to obtain $Pe_s$ values for active particles at which they start going outside of the apparatus. The criteria were escaping 5\% of particles after 500000 steps with a timestep of dt = 0.0001. This was the main component to have demixing for particles/apparatus with different properties. After obtaining the data from simulations, we proceeded with different machine learning algorithms to model active particle motility (P\`eclet), each algorithm is considered in the following sections.

\subsection{Machine Learning Models}
\label{sec:ML}
\subsubsection{Artificial Neural Network}
\label{sec:ANN}

In this section, we use ANN to model our systems. Typical workflow for modeling using ANN involves several steps, starting with collecting data and preprocessing it, preparing the model, training it, and evaluating the performance of the model. For collecting data, as stated earlier, 2500 relevant data points from a set of 75000 simulations were used. Input parameters for the data was $r_s$, $\omega$, $d_{mono}$, $l_{pore}$, and $k_BT$ with target values being $Pe_s$. Then we preprocessed our data using the Scikit-learn package \cite{pedregosa2011scikit}, this step transforms the data into a format that is more convenient to train ANN.

\begin{figure}
	\centering
	\includegraphics[width=0.6\linewidth]{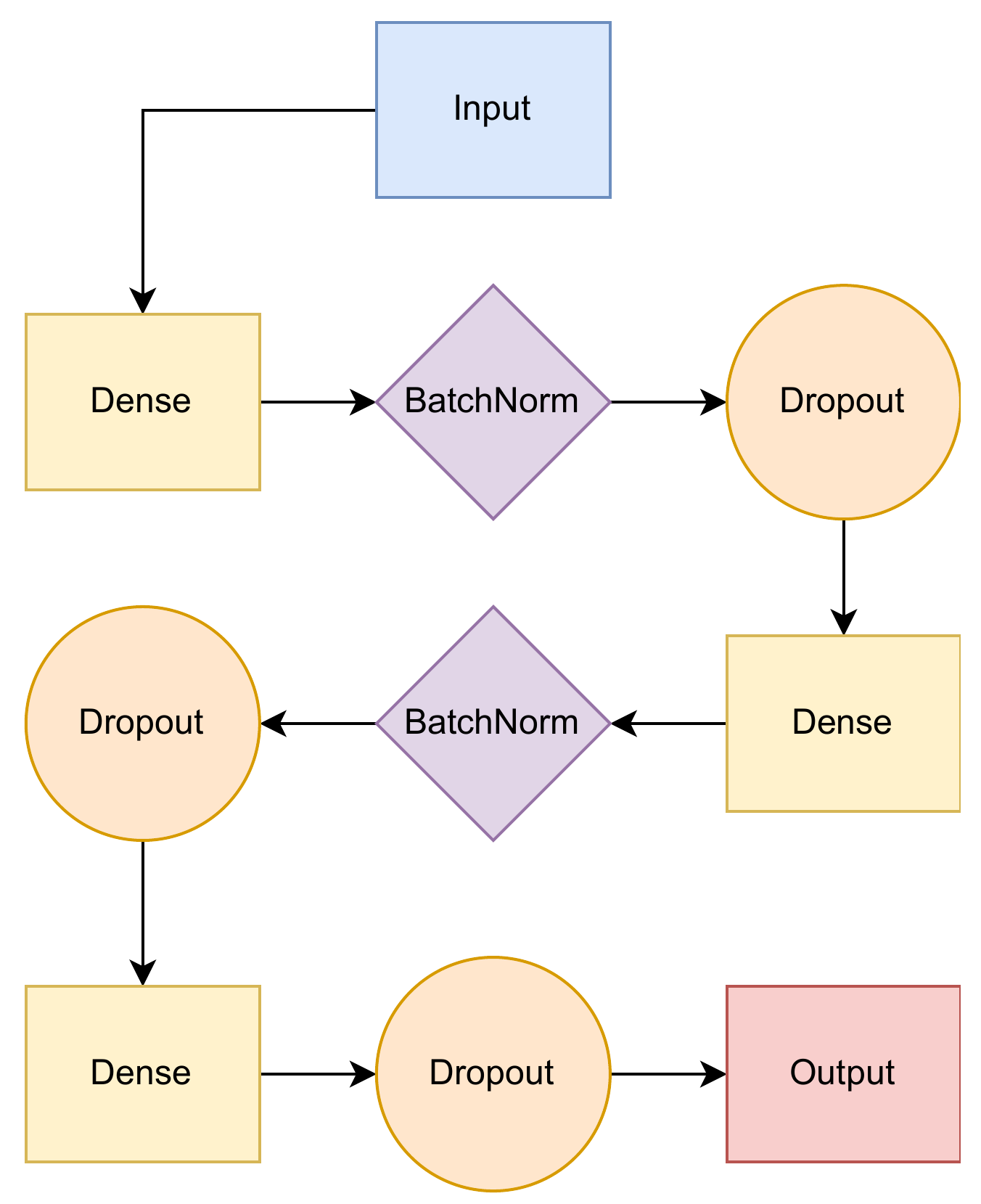}
	\caption{ANN architecture containing the input, dense layer with 512 neurons, a bath normalization layer, a dropout layer, dense layer with 128 neurons, a bath normalization layer, a dropout layer,  dense layer with 64 neurons, a dropout layer, and the output layer} 
	\label{fig:nnarch}
\end{figure}

Using Keras from TensorFlow, we created a dense (fully connected network), meaning that each of the neurons in a layer is connected to every neuron in the preceding and subsequent layers, this interconnectivity would make the ANN capable of modeling complex and non-linear relationships between data. The architecture of our ANN which was designed for regression is comprised of (Fig \ref{fig:nnarch}): an input layer, a dense layer where each unit is connected to every feature in the input data, and a Rectified Linear Unit (ReLU) activation function. This was followed by a Batch Normalization Layer, we included this layer to normalize the activation of the previous layer to enhance the training stability and convergence speed. A Dropout Layer was added to "drop out" (set to zero) a fraction of the neurons from the previous layer at random while training the network. Using this we prevent the model from relying heavily on a neuron or a feature. Thus, making the network more general and not specialized and overly complex, in simple words prevents overfitting. This was succeeded with another dense layer and a ReLU activation function. This layer processes information from the input and learns to recognize complex patterns in the data. This layer also was followed by Batch Normalization and Dropout layers. A subsequent hidden layer with ReLU activation was included, followed by a dropout layer. Finally, we have the output layer which contains a single unit, appropriate for regression tasks where the goal is to predict a continuous value.

%	Hyperparameter Tuning?
Following defining the ANN architecture, hyperparameters of the model were optimized to enhance the predictive performance, optimized hyperparameters are listed in Table \ref{tab:models-params}. We trained the model utilizing a custom Adam optimizer with the determined learning rate. The loss function we chose was MSE. We also utilized early stopping while training the model, this technique monitors MSE on a validation set and when the model has no improvement over a defined number of epochs, training stops. This prevents learning noise and aids generalization to unseen data. 

After completion of training the ANN model, we assessed it on the data that we used to train the model and also the test set. We also used another external test set, which was not used or seen by the neural net, and we assessed it after getting the final model to see its predictive ability. We employed R-squared ($R^2$), Mean Absolute Error (MAE), Root Mean Squared Error (MSE), and Mean Absolute Percentage Error (MAPE) as evaluation metrics, these showcase the predictive capabilities of the model, and are defined in the following:
\begin{align}
	R^2 &= 1 - \frac{{\sum\limits_{i=1}^{N}\left(y_i - \hat{y}_i\right)^2}}{{\sum\limits_{i=1}^{N}\left(y_i - \bar{y}\right)^2}}\\[2pt]
	MAE &= \frac{{\sum\limits_{i=1}^{N}\left|y_i - \hat{y}_i\right|}}{{N}}\\[2pt]
	RMSE &= \sqrt{\frac{{\sum\limits_{i=1}^{N}\left(y_i - \hat{y}_i\right)^2}}{{N}}}\\[2pt]
	MAPE &= \frac{{\sum\limits_{i=1}^{N}\left|\frac{{y_i - \hat{y}_i}}{{y_i}}\right|}}{{N}} \times 100\%
\end{align}
$N$ is the total number of data points that are getting evaluated, $y_i$ , $\hat{y}_i$, and $bar{y}$ are the real, predicted, and mean values of the entity being evaluated, respectively. For ANN model evaluation metrics are presented in Table \ref{tab:model-eval-metrics}. 

\subsubsection{Support Vector Regression}
\label{sec:SVM}

In this section, we discuss the utilization of Support Vector Regression (SVR) to model our systems. The modeling steps for SVR are similar to those for ANN: data collection, data preprocessing, model preparation, model training, and model evaluation performance. Data collection and preprocessing are the same as before. To implement the SVR model, we used the scikit-learn library, specifically Epsilon-Support Vector Regression. 

We used the Radial Basis Function (RBF) kernel in our SVR model. It is defined by the formula: $K(x, x') = e^{-\gamma \cdot \lVert x - x' \rVert^2}$. Here, $x$ and $x'$ are the input vectors, $||x - x'||^2$ is the squared distance between them, and gamma is a parameter that controls the width of the Gaussian distribution. The RBF kernel can effectively capture complex, non-linear relationships in the data by transforming the data using the preceding formula. Gamma is the hyperparameter that needs to be tuned for optimal model performance.

Other hyperparameters for implementing this SVR model include C (regularization parameter) and epsilon (insensitivity parameter). Hyperparameter C controls how the model treats outliers. A low value of C implies that more outliers are allowed, while a high value of C implies that fewer outliers are tolerated. Hyperparameter epsilon is given by $L(y, f(x), \epsilon) = \max(0, |y - f(x)| - \epsilon)$, where $y$ represents the true target value, $f(x)$ is the predicted target value and $\epsilon$ denotes the width of the insensitive zone, controlling the tolerated error. Optimum hyperparameters and evaluation metrics for our SVR model are listed in Tables \ref{tab:models-params}, \ref{tab:model-eval-metrics}, respectively.

\subsubsection{Kernel Ridge Regression}
\label{kernel}

Similar to the SVR, Kernel Ridge Regression (KRR) transforms the input data into a higher-dimensional space using a nonlinear kernel function. This makes it possible for the algorithm to learn complex relationships between the input data and the outputs. In this section, we will introduce and use KRR to model our systems.

The objective function in KRR is defined by the equation:
\begin{align}
	\hat{\boldsymbol{\alpha}} = \arg\underset{\boldsymbol{\alpha}\in\mathbb{R}^N}{\min} \ \|\mathbf{y} - \mathbf{K}\boldsymbol{\alpha}\|_2^2 + \lambda \boldsymbol{\alpha}^T \mathbf{K} \boldsymbol{\alpha}.
\end{align}
KRR is designed to fit the training data while controlling overfitting through regularization. The objective is to find the vector $\hat{\boldsymbol{\alpha}}$ that minimizes the squared Euclidean norm between $\mathbf{y}$ and $\mathbf{K}\boldsymbol{\alpha}$ plus a regularization term ($\|\mathbf{y} - \mathbf{K}\boldsymbol{\alpha}\|_2^2 + \lambda \boldsymbol{\alpha}^T \mathbf{K} \boldsymbol{\alpha}$). Here, $\boldsymbol{\alpha}$ is the coefficients (weights) assigned to each training sample in the kernelized feature space which is getting optimized. $\mathbf{y}$ is the real (target) value. $\mathbf{K}$ is the kernel matrix where each element is the result of applying a kernel function to the pair of input samples and $\mathbf{K}\boldsymbol{\alpha}$ is the transformed input data. The squared Euclidean norm between $\mathbf{y}$ and $\mathbf{K}\boldsymbol{\alpha}$ measures the deviation between real and predicted values. The regularization term is $\lambda \boldsymbol{\alpha}^T \mathbf{K} \boldsymbol{\alpha}$, where $\lambda$ is a constant parameter that specifies the strength of the regularization to prevent overfitting.

We implemented the KRR using the scikit-learn library and performed a grid search to obtain the optimal values of the hyperparameters. The hyperparameters included in the optimization were regularization strength, kernel type (Linear, polynomial, RBF, and sigmoid), gamma, which denotes the kernel coefficient (the width of the kernel) that is specific to each kernel being used, and the polynomial degree for the polynomial kernel (ignored by other kernels). Best performing hyperparameters obtained for the KRR model and the evaluation metrics for the KRR model are listed in Tables \ref{tab:models-params}, \ref{tab:model-eval-metrics}, respectively.

\subsubsection{Gaussian Process Regression}
\label{gaussian}
Gaussian Process Regression (GPR) is a probabilistic modeling technique used to capture non-linear relationships in data to perform regression tasks. Unlike conventional methods, GPR treats the target variable as a distribution over functions, rather than yielding single-point estimates. This allows for making predictions with associated uncertainty.

To perform a regression task using GPR, initially, we establish a prior belief, characterized by a mean function $m(x)$ and a covariance function (kernel) $(k(x, x')$, before any data is observed. This represents the initial understanding of the function space. Combining the prior belief with the observed data results in a posterior distribution (Bayes' rule) which is Gaussian. The posterior is a refined understanding of the function space. This procedure is iterative, and with each new data point, the model's accuracy increases.

%	The choice of kernel plays a crucial role in shaping the assumed smoothness and patterns in the function. For instance, the Radial Basis Function (RBF) kernel, represented by \(k(x, x') = \exp\left(-\frac{\|x - x'\|^2}{2l^2}\right)\), is commonly employed for capturing smooth, continuous variations in the data.

During training, we compute the posterior mean $\mu$ and covariance $\Sigma$ using the following equations:
\begin{equation}
	\begin{aligned}[b]
		\mu(X_*, X, \mathbf{y}) = K(X_*, X) [K(X, X) + \sigma_n^2 I]^{-1} \mathbf{y}
	\end{aligned}
\end{equation}
\begin{equation}
	\begin{aligned}[b]
		& \Sigma(X_*, X) = K(X_*, X_*)\\
		& - K(X_*, X) [K(X, X) + \sigma_n^2 I]^{-1} K(X, X_*)
	\end{aligned}
\end{equation}
$K$ is the kernel matrix with $K_{ij} = k(x_i, x_j)$. This captures the pairwise similarities between input points. $X_*$ represents new input points, $\mathbf{y}$ is the vector of observed values, $\sigma_n^2$ is the variance of the noise, and $I$ is the identity matrix. When making predictions for new inputs, the posterior mean $\mu$ is the predicted value, while covariance $\Sigma$ indicates the associated uncertainty.

We used the scikit-learn library to implement GPR for modeling our systems. Specifically, we used the method therein called GaussianProcessRegressor with a custom kernel function which is a combination of the product of a constant kernel (C) by an RBF kernel plus a white kernel to account for noise in data plus a "DotProduct" kernel with power four to act as a four-degree polynomial. Both the RBF kernel and DotProduct kernel (with power 4) account for the non-linear relationship between the input data and the targets. We used default hyperparameters for our GPR, listed in Table \ref{tab:models-params}, the reason being that all of these hyperparameters are optimized during the training by the implementation of GPR in the scikit-learn library. Evaluation metrics for our GPR model are also listed in Table \ref{tab:model-eval-metrics}.

\begin{table}[h]
	\caption{ML models hyperparameters}
	\label{tab:models-params}
	\small % Set small font size
	\renewcommand{\arraystretch}{1.3} % Adjust space between rows as needed
	\begin{tabular}{p{0.15\columnwidth} p{0.4\columnwidth} p{0.35\columnwidth}}
		\hline
		Model & Hyperparameter & Details\\ 
		\hline
		ANN & optimizer & Adam \\ 
		& max\_iter & 10000 \\
		& early stopping patience & 2000 \\
		& learning\_rate & 0.001 \\
		& batch\_size & 512 \\
		& hidden neurons & 512, 128, 64 \\
		& activations & relu \\
		& loss & mse \\
		\hline
		SVR & kernel& rbf \\
		& C & 100000 \\
		& epsilon & 0.9 \\
		& gamma & 0.05 \\
		\hline
		KRR & kernel& polynomial\\
		& degree & 4 \\
		& $\lambda$ &0.5 \\
		& gamma & 0.2 \\
		\hline
		GPR & kernel& Constant*rbf+ WhiteKernel+ $\text{DotProduct}^4$\\
		& alpha & 0.1\\
		%		=1**2 * RBF(length_scale=1) + WhiteKernel(noise_level=1) + DotProduct(sigma_0=1) ** 4,
		& n\_restarts\_optimizer & 10\\
		
		\hline
	\end{tabular}
\end{table}

\subsubsection{Comparing the ML methods}

Here, we compare the overall performance of all models using evaluation metrics listed in Table \ref{tab:model-eval-metrics}. The $R^2$ values for all models are high, approaching unity, across all three datasets, suggesting a strong correlation between the variables and the target values. We also note that a high $R^2$ value for the training set would indicate overfitting, considering the table, $R^2$ values for test and external test sets are high and near the values of $R^2$ for their respective training set, this would mitigate the overfitting problem, thus, validating the generalizability of our models. The MAPE, MAE, and RMSE evaluation metrics too, further showcase consistent values across all three datasets, with minor variations, this too denotes the generalizability and consistency in the predictive power of our models.

\begin{table}[h]
	\centering
	\caption{Evaluation Metrics for Different Models}
	\label{tab:model-eval-metrics}
	\small % reduce font size
	\setlength{\tabcolsep}{4pt} % adjust the space between columns
	\renewcommand{\arraystretch}{1.1} % adjust row height as suitable
	\begin{tabular}{llcccc}
		\hline
		\textbf{Model} & \textbf{Eval.} & \textbf{R$^2$} & \textbf{MAPE} & \textbf{MAE} & \textbf{RMSE} \\ 
		\hline
		NN & Train & 0.9997 & 1.9539 & 1.1007 & 1.5084 \\
		& Test & 0.9997 & 2.1825 & 1.2077 & 1.5331 \\
		& Ext. & 0.9997 & 1.9725 & 1.2286 & 1.6334 \\ 
		\hline
		SVR & Train & 0.9996 & 1.4968 & 1.1768 & 1.8964 \\ 
		& Test & 0.9995 & 1.6141 & 1.3136 & 2.0397 \\ 
		& Ext. & 0.9993 & 1.7029 & 1.4192 & 2.4055 \\ 
		\hline
		KRR & Train & 0.9994 & 2.1512 & 1.5838 & 2.4108 \\ 
		& Test & 0.9992 & 2.0918 & 1.6233 & 2.4612 \\ 
		& Ext. & 0.9991 & 2.2392 & 1.7889 & 2.7267 \\ 
		\hline
		GPR & Train & 0.9995 & 2.7235 & 1.5265 & 2.2082 \\ 
		& Test & 0.9994 & 2.6323 & 1.5563 & 2.2458 \\ 
		& Ext. & 0.9993 & 2.7110 & 1.6136 & 2.4577 \\ 
		\hline
	\end{tabular}
\end{table}

\begin{figure}[h]
	\centering
	\subfloat[NN]{
		\centering
		\includegraphics[width=0.47\linewidth]{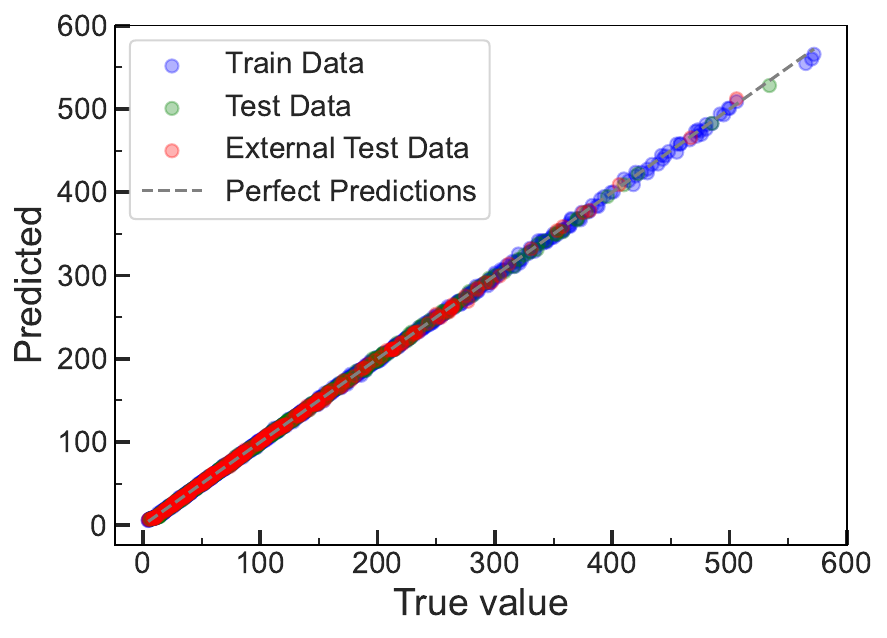}
		\label{fig:pred-obs1}}
	\hspace{-4mm}
	\subfloat[SVR]{
		\centering
		\includegraphics[width=0.47\linewidth]{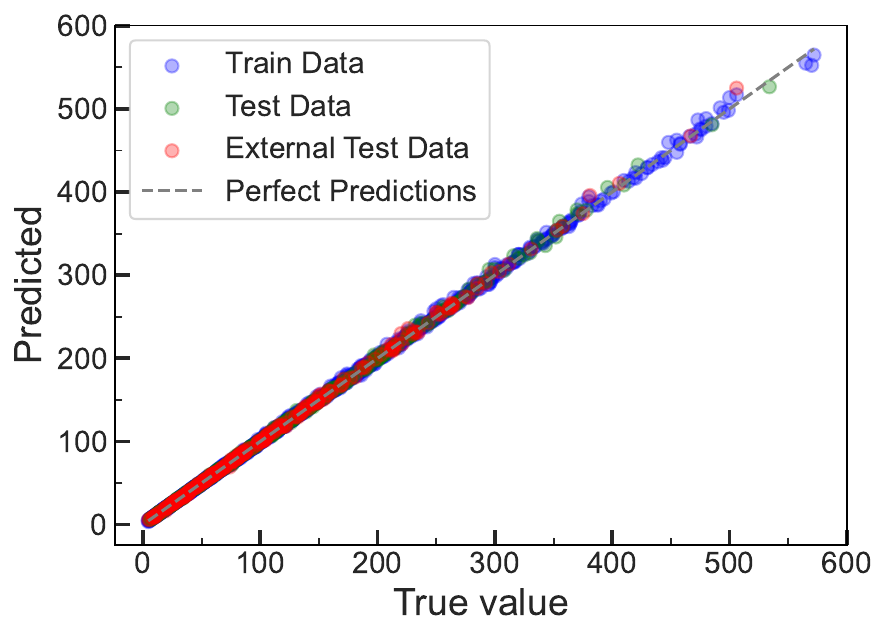}
		\label{fig:pred-obs2}}
	\hspace{-4mm}
	\subfloat[KRR]{
		\centering
		\includegraphics[width=0.47\linewidth]{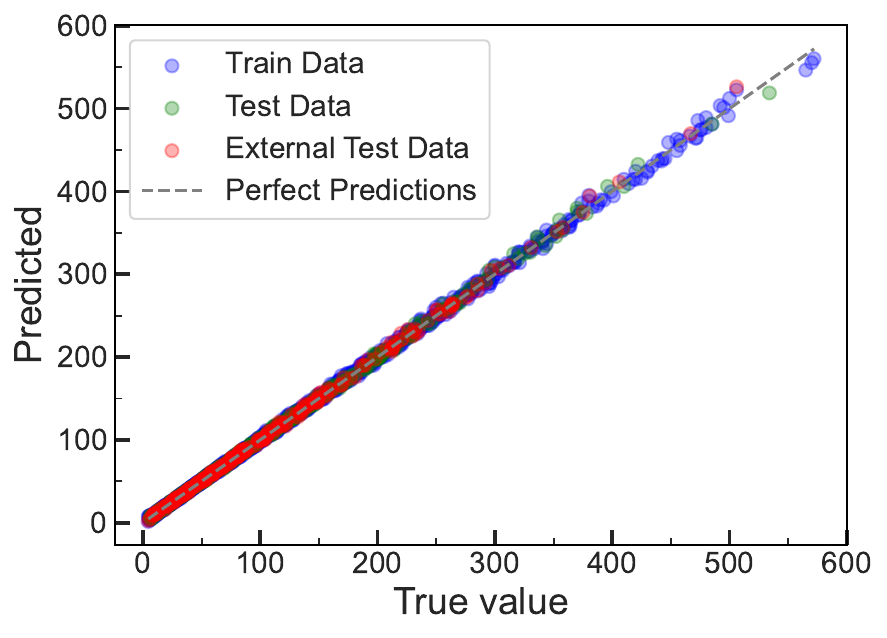}
		\label{fig:pred-obs3}}
	\hspace{-4mm}
	\subfloat[GPR]{
		\centering
		\includegraphics[width=0.47\linewidth]{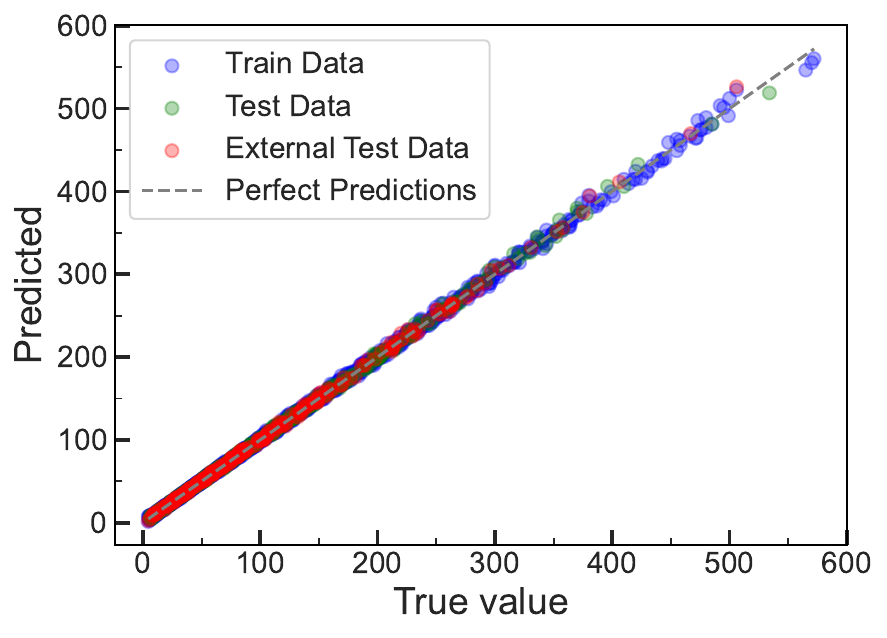}
		\label{fig:pred-obs4}}
	\hspace{-4mm}
	\caption{Plots of prediction vs observed value for our models: a)NN, b)SVR, c)KRR, and d)GPR.}
	\label{fig:pred-obs}
\end{figure}

We also included Fig.~\ref{fig:pred-obs}, which illustrates the predicted values plotted against the observed values for each ML method we used, it includes the data for all three datasets -- training, testing, and external validation -- and also a guideline to show the perfect predictions (benchmark). Each plot shows how closely predictions from each of the ML models match up to the actual target values across the datasets. 

Overall, based on the evaluation metrics and also plots of predicted vs observed values, the predictive power of the ML methods from the highest performing to the lowest one -- with a minimal variation -- is: ANN, SVR, GPR, and SVR respectively. 

\subsubsection{Model Interpretation}
All the models we used, as well as many ML methods, present challenges to rationalizing and interpreting the results of predictions, this is due to their intrinsic black-box nature. To overcome this shortcoming we will use the SHapley Additive exPlanations (SHAP) method, utilizing SHAP we can measure the contribution of each feature on the model's predictions \cite{lundberg2017unified}.

\begin{figure}
	\centering
	\includegraphics[width=0.95\linewidth]{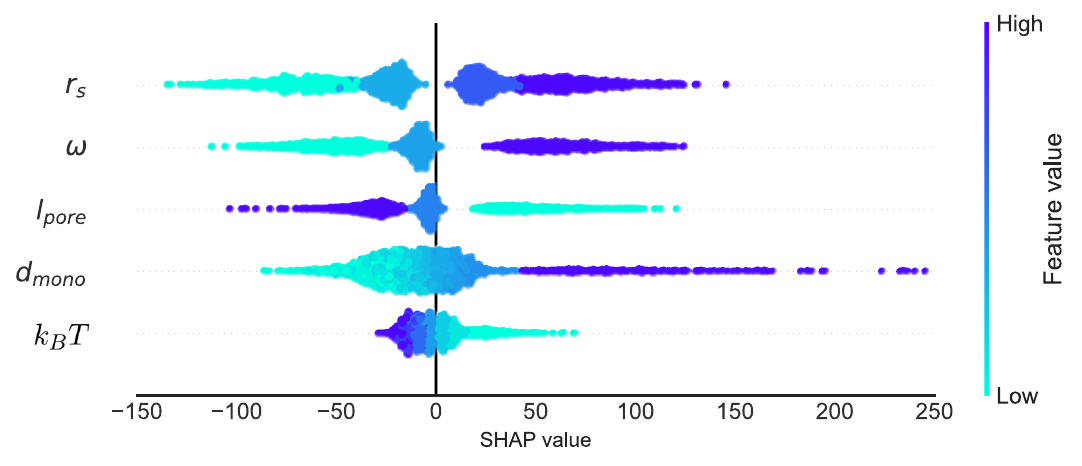}
	\caption{Feature importance using SHAP analysis. The x-axis represents the SHAP values and the y-axis represents the feature names. Feature values change from cyan to blue as they increase from low to high. The most important feature is placed at the top. Here, a positive (negative) SHAP value for a feature means that the prediction increases (decreases) with an increase (decreases) in the given feature value.}
	\label{fig:shap_directionality}
\end{figure}

\begin{figure*}[htb!]
	\centering
	\includegraphics[width=0.90\linewidth]{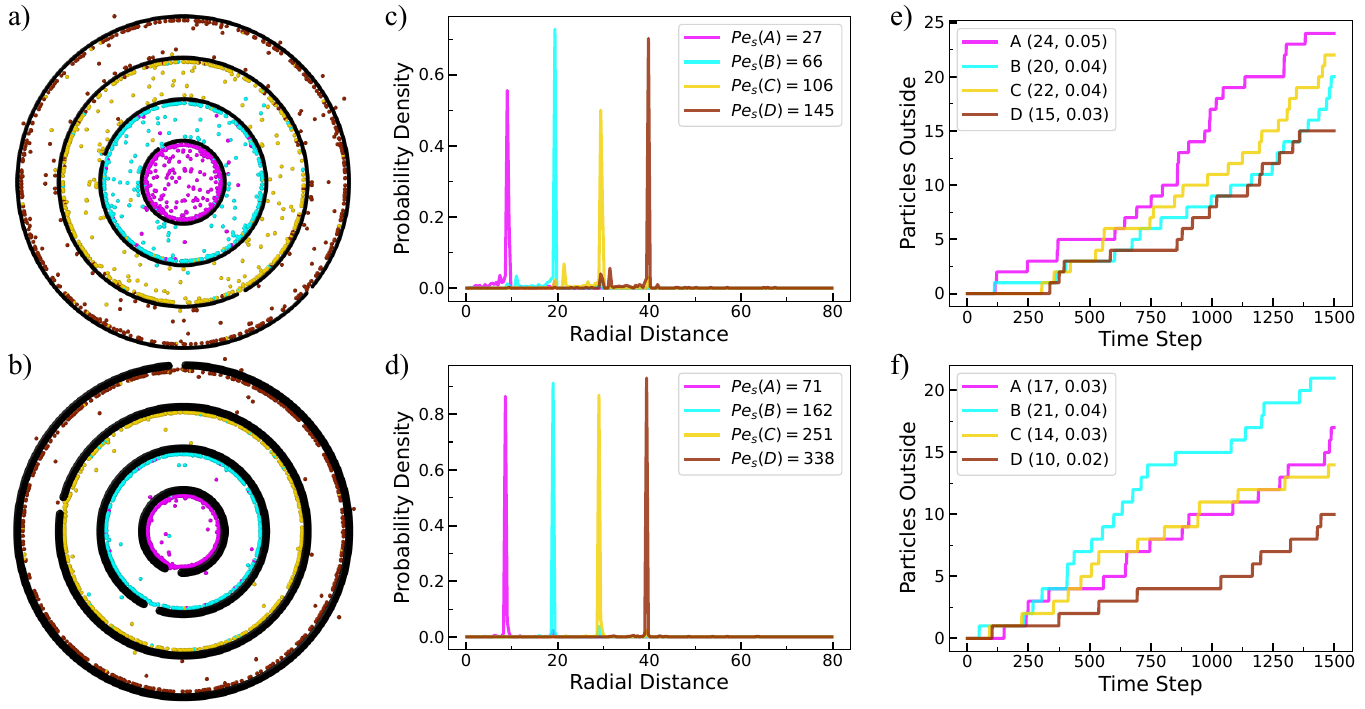}
	\caption{Last snapshots of the simulations, radial probability distributions, and also number of particles going outside of their respective confinements. a, c, e for system with $\omega$=1.0, $k_BT$=1.0, $l_{pore}$=2.0, $d_{mono}$=1.0, and $Pe_s$: 27.28, 66.11, 105.65, 145.47. b, d, f for systems with $\omega$=3.0, $k_BT$=1.0, $l_{pore}$=4.0, $d_{mono}$=2.0, and $Pe_s$: 70.57, 162.37, 250.83, 338.16. Colors indicate particle types with specific P\'eclet values that are predicted with ANN. In the legend for e and f, the first number indicates the number of particles that escaped their confinement, and the second value is their ratio (same color).}
	\label{fig:systems}
\end{figure*}

In Figure \ref{fig:shap_directionality}, $x$ coordinates show the SHAP values and the y coordinate depicts the feature names, there is also color coding for the values of each feature from low to high when going from cyan to blue. This plot also shows the importance of features, with the most important one placed at the top. We also note that a positive (negative) SHAP value for one feature means the prediction, namely the P\`eclet value, increases (decreases) with the given feature value. Feature importance from the figure is: $r_s > \omega > l_{pore} > d_{mono} > k_B T$. Further, from the color coding in the figure, for radius ($r_s$), increasing the feature values increases P\`ecelt values. This is in line with the apparatus design and configuration of our systems, higher radius ($r_s$) values would mean a higher linear velocity for the monomers of the apparatus while all other parameters including angular velocity ($\omega$) are kept constant, this translates into higher P\`eclet values needed for active particles to escape outside of the apparatus. This is the same for the angular velocity of the apparatus, an increase in this feature would lead to an increase in the velocity needed for particles to pass through the pore of the apparatus, as it is shown from color coding for $\omega$. Higher monomer diameter $d_{mono}$ also shows increased SHAP values, considering that when all parameters are kept constant, a higher $d_{mono}$ would mean smaller pore length and bigger contact area for the swimmers to pass through the pore. For pore length $l_{pore}$ this means that a higher value of the feature would decrease the P\`eclet needed for active particles to go outside the apparatus, as evident from the figure. Finally, as it can be seen from the figure, increasing $k_BT$ decreases the SHAP values, it is intuitive that increasing $k_BT$ values would mean a higher activity of the swimmers, and the P\`eclet values needed to escape get smaller.

\subsubsection{Demixing using ML}

Here, we showcase how our ML models define our systems capable of demixing. We consider our best-performing model, namely ANN. Figures \ref{fig:systems}a and \ref{fig:systems}b show the last snapshots of simulations of two separate systems in which we modeled the system using our ANN model. This figure also includes the radial probability distributions of the particles for their respective systems, Figures \ref{fig:systems}c and \ref{fig:systems}d. Further, we show the number of particles that escape their respective confinements and the ratio of escaped particles to the total particles of the respective type in Figures \ref{fig:systems}e and \ref{fig:systems}f. 
Figure \ref{fig:systems}a is the last snapshot of the system with $\omega$=1.0, $k_BT$=1.0, $l_{pore}$=2.0, $d_{mono}$=1.0. $Pe_s$ values predicted for this system using our GPR model are $Pe_s$: 27.28, 66.11, 105.65, 145.47. We used these parameters and $Pe_s$ values to simulate the system, and from the figure, we see the onset of demixing. Figure \ref{fig:systems}c shows the Radial Probability Distribution for this system, from this, we see that most of the particles remain in between their respective confinements. From Figure \ref{fig:systems}e, 95\% of particles of type A remain below the radius 10 and only 5\% of them go outside. For other types of particles, namely B, C, and D, less than 5\% of each goes above their respective confinement radii, namely 20, 30, and 40 respectively. 

Other system with $\omega$=3.0, $k_BT$=1.0, $l_{pore}$=4.0, $d_{mono}$=2.0 and predicted values of $Pe_s$: 70.57, 162.37, 250.83 and 338.16 showcase a similar pattern (see Figures \ref{fig:systems}b,\ref{fig:systems}d,\ref{fig:systems}f), showing effectiveness of our ANN to predict P\'eclet values for demixing.

\FloatBarrier
%%--------------------Discussions----------------------------------------------------------
\section{Discussion}
\label{sec:Discussion}

The trapping phenomena of active particles, as observed in various experimental and numerical studies, have been instrumental in advancing our understanding of how geometrically designed boundaries can be leveraged for effective substance purification and microorganism control. These studies have demonstrated the potential of active \cite{Kaiser:PRL2012,ribeiro2020trapping,kumar2019trapping} and chiral active \cite{ai2015chirality,chen2015sorting,Ai:SoftMatter2018,Ai_2023,Levis:PRR2019,Barois:PRL2020} particles to navigate through geometrically designed boundaries, offering promising applications in the purification of substances by selectively eliminating unwanted microorganisms. A notable aspect of these studies is the exploration of narrow escape times of active particles from a circular domain, which was achieved using numerical simulation \cite{paoluzzi2020narrow}. In addition, trapping particles using an array of funnel-shaped barriers with wide and narrow openings was developed to concentrate bacteria (E. coli) suspensions with motility on the side with narrow openings \cite{galajda2007wall}. Another study investigated the impact of spatially periodic potentials on trapping and sorting motile active particles \cite{ribeiro2020trapping}. Furthermore, a study proposed a method that incorporates an acoustofluidic setup for selecting particles based on their motility, where particles with high motility would escape from the acoustic trap \cite{misko2023selecting}. This method was demonstrated using both simulations and experiments with Janus particles and human sperm, proposing that this method could be used to select highly motile sperm for medically assisted reproduction. All of the previous research focused on entrapment and sorting active particles, however, one key challenge that remained unexplored was sorting microorganisms with various propulsions with a high selectivity.

In the present study, we have developed a novel mechanism to sort active particles based on their motility values, specifically the P\`eclet number, utilizing a combination of Brownian dynamics simulations and machine learning methods. Our proposed mechanism is based on a circular spinning apparatus with pores allowing active particles with certain P\`eclet numbers to pass through it. Initially, we devised a system with a configuration that would demix and separate active particles with four different motilities. To better understand the system configurations required to separate motile particles, considering their complex and non-equilibrium dynamics, the utilization of machine learning methods was necessary. 

Using Brownian dynamics we conducted an extensive number of simulations to generate a dataset to be used by the machine learning methods to capture the complex dynamics of active particles. After gathering the data, four machine learning methods, namely: artificial neural network, support vector regression, kernel ridge regression, and Gaussian process regression were developed, with the best performing one being the artificial neural network. Further, we showcased the onset of demixing and sorting based on the P\`eclet numbers predicted by the neural network, simulation snapshots, plots of radial probability distribution, and also plots of the escape number of particles from the apparatus were included as visual and numerical supplements to our findings. We also employed the SHAP method to consider the importance of system configurations and to interpret the model's predictions, this revealed that the radius ($r_s$) and angular velocity ($\omega$) of the apparatus, the monomer diameter ($d_{mono}$), and the pore length ($l_{pore}$) significantly influenced the P\`eclet numbers required for active particles to escape the apparatus, while $k_BT$ values, indicating the noise part of the motion, resulted in lower required P\`eclet numbers and had the lowest impact.

Although our system was confined to two dimensions, such a mechanism could easily be implemented in 3D experimental setups. However, further research is needed to explore the effectiveness of this mechanism with different motility parameters and in real-world systems. Further, with significant advancements in microfluidic device fabrication techniques \cite{niculescu2021fabrication, scott2021fabrication}, and the introduction of 3D printing technologies, rapid and single-step production of intricate microfluidic devices is not only more accessible but has also expanded the design possibilities \cite{Duong2019microfluidic3d, waheed20163d}, \cite{Amin2016microfluidic3d}. Such devices show great potential as advanced systems for manipulating and immobilizing small organisms. Therefore, we believe that our proposed mechanism and apparatus can be designed and fabricated with an autonomous mechanism, opening up possibilities for applications in various fields such as physics, biology, and drug delivery.

\begin{acknowledgements} 
This work was supported by the Iran National Science Foundation (INSF) via Grant No. 4003065. Computational resources were provided by the Center for High-Performance Computing (SARMAD) at Shahid Beheshti University, Tehran. We acknowledge useful discussions with R. K. Bowles. 
\end{acknowledgements}

\bibliography{ref}

\end{document}